\documentclass[journal,final,a4paper,onecolumn,10]{IEEEtran}

\ifCLASSINFOpdf
\else
\fi

\usepackage{amsmath,graphicx}
\usepackage{amssymb,epsfig,color}

\newcommand{\vm}[1]{\ensuremath{\bm{#1}}}% vector or matrix

\renewcommand{\exp}[1]{\ensuremath{\text{e}^{#1}}}% exponential
\renewcommand{\Re}[1]{\ensuremath{\text{Re}\!\left\{#1\right\}}}% real part
\renewcommand{\Im}[1]{\ensuremath{\text{Im}\!\left\{#1\right\}}}% imaginary part
\newcommand{\cell}{c}
\newcommand{\cellIn}{\tilde{c}}
\newcommand{\cellt}{\mathbf{\cell}^{(t)}}
\newcommand{\celltLast}{\mathbf{\cell}^{(t-1)}}
\newcommand{\cellInt}{\mathbf{\cellIn}^{(t)}}

\newcommand{\xt}{\mathbf{x}^{(t)}}

\newcommand{\y}{\mathbf{y}}

\newcommand{\igate}{\mathbf{i}^{(t)}}
\newcommand{\fgate}{\mathbf{f}^{(t)}}
\newcommand{\ogate}{\mathbf{o}^{(t)}}

\newcommand{\hidt}{\mathbf{h}^{(t)}}

\newcommand{\hidtLast}{\mathbf{h}^{(t-1)}}

\newcommand{\bi}{\mathbf{b}_{i}}
\newcommand{\bfgate}{\mathbf{b}_{f}}
\newcommand{\bo}{\mathbf{b}_{o}}
\newcommand{\bcin}{\mathbf{b}_{c}}

\newcommand{\Wix}{\mathbf{W}_{xi}}

\newcommand{\Wih}{\mathbf{W}_{hi}}
\newcommand{\Wfx}{\mathbf{W}_{xf}}

\newcommand{\Wfh}{\mathbf{W}_{hf}}
\newcommand{\Wox}{\mathbf{W}_{xo}}

\newcommand{\Woh}{\mathbf{W}_{ho}}
\newcommand{\Wcx}{\mathbf{W}_{xc}}

\newcommand{\Wch}{\mathbf{W}_{hc}}

\newcommand{\Sih}{s_{hi}}
\newcommand{\Soh}{s_{ho}}
\newcommand{\Sfh}{s_{hf}}
\newcommand{\Sch}{s_{hc}}

\usepackage{bm}
\usepackage[utf8]{inputenc}
\usepackage{url}
\usepackage[table]{xcolor}
\usepackage{multirow}
\usepackage{makecell}

\usepackage{caption}
\usepackage{subfig}
\usepackage{fixltx2e}      %required for \textsubscript{}

\newcommand\ssf[1]{{\small\textsf{#1}}}
\newcommand\fsf[1]{{\footnotesize\textsf{#1}}}

\begin{document}

\title{Resource-Efficient Speech Mask Estimation for Multi-Channel Speech Enhancement}


%\author{Lukas~Pfeifenberger, Matthias~Z\"ohrer, Wolfgang Roth, G\"unther Schindler, Holger Fr\"oning and Franz Pernkopf,~\IEEEmembership{Senior Member,~IEEE}
\author{Lukas~Pfeifenberger, Matthias~Z\"ohrer, G\"unther Schindler, Wolfgang Roth, Holger Fr\"oning and Franz Pernkopf,~\IEEEmembership{Senior Member,~IEEE}
%\thanks{L.~Pfeifenberger, M.~Z\"ohrer, W. Roth and F. Pernkopf are with the Intelligent Systems Group at the Signal Processing and Speech Communication Laboratory, Graz University of Technology, Graz, Austria. G. Schindler and H. Fr\"oning are with the Institute of Computer Engineering, Ruperts Karls University, Heidelberg, Germany.}
\thanks{L.~Pfeifenberger, M.~Z\"ohrer, W. Roth and F. Pernkopf are with the Intelligent Systems Group at the Signal Processing and Speech Communication Laboratory, Graz University of Technology, Graz, Austria. }
\thanks{G. Schindler and H. Fr\"oning are with the Institute of Computer Engineering, Ruperts Karls University, Heidelberg, Germany.}
\thanks{This work was supported by the Austrian Science Fund (FWF) under the project number I2706-N31. We acknowledge Ognios GmbH for supporting Lukas Pfeifenberger. Furthermore, we acknowledge NVIDIA for providing GPU computing resources.}}

%\IEEEpeerreviewmaketitle

\maketitle

\begin{abstract}
While machine learning techniques are traditionally resource intensive, we are currently witnessing an increased interest in hardware and energy efficient approaches. This need for resource-efficient machine learning is primarily driven by the demand for embedded systems and their usage in ubiquitous computing and IoT applications. In this article, we provide a resource-efficient approach for multi-channel speech enhancement based on Deep Neural Networks (DNNs). In particular, we use reduced-precision DNNs for estimating a speech mask from noisy, multi-channel microphone observations. This speech mask is used to obtain either the \emph{Minimum Variance Distortionless Response} (MVDR) or \emph{Generalized Eigenvalue} (GEV) beamformer. In the extreme case of binary weights and reduced precision activations, a significant reduction of execution time and memory footprint is possible while still obtaining an audio quality almost on par to single-precision DNNs and a slightly larger \emph{Word Error Rate} (WER) for single speaker scenarios using the WSJ0 speech corpus. 
\end{abstract}

%\includegraphics{figs/grabs.pdf}

%\item DNNs with reduced-precision weights are used for speech mask estimation %in multi-channel speech enhancement (MCSE).
%\item Computational efficiency of the MCSE system is analysed and performance %metrics of binary matrix multiplications on ARM CPUs and NVIDIA GPUs are %provided.
%\item A tradeoff analysis (separation performance vs. computational und %representational efficiency) for speech mask estimation using deep neural %networks (DNNs) is performed.
%\item Word error rate and SNR scores for MCSE using reduced-precision DNNs are %reported in experiments. 

\begin{IEEEkeywords}
Resource-efficient machine learning, multi-channel speech enhancement, speech mask estimation, binary neural networks 
\end{IEEEkeywords}

\IEEEpeerreviewmaketitle

\section{Introduction}
\label{sec:introduction}
Deep Neural Networks (DNNs), the workhorse of  speech-based user interaction systems, prove particularly effective when big amounts of data and plenty of computing resources are available. However, in many real-world applications the limited computing infrastructure, latency and the power constraints during the operation phase effectively suspend most of the current resource-hungry DNN approaches. Therefore, there are several key challenges which have to be jointly considered to facilitate the usage of DNNs when it comes to edge-computing implementations:
\begin{itemize}
\item Efficient representation: The model complexity measured by the number of model parameters should match the limited resources of the computing hardware, in particular regarding memory footprint.
\item Computational efficiency: The model should be computationally efficient during inference, exploiting the available hardware optimally with respect to time and energy. Power constraints are key for embedded systems, as the device lifetime for a given battery charge needs to be maximized.
\item Prediction quality: The focus is usually on optimizing the prediction quality of the models. For embedded devices, model complexity versus prediction quality trade-offs must be considered to achieve good prediction performance while simultaneously reducing computational complexity and memory requirements.
\end{itemize}    

DNN models use GPUs to enable efficient processing, where single precision floating-point numbers are common for parameter representation and arithmetic operations. To facilitate deep models in today's consumer electronics, the model usually has to be scaled down to be implemented efficiently on embedded or low power systems. Most research emphasizes one of the following two techniques: (i) reduce model size in terms of number of weights and/or neurons~\cite{Han2015,Han2016,Gordon:2018,Wu:2018,Howard:2017,Zoph:2018}, or (ii) reduce arithmetic precision of parameters and/or computational units~\cite{Courbariaux2015b,dorefa,OttLZLB16,per18}. Evidently, these two basic techniques are almost ``orthogonal directions'' towards efficiency in DNNs, and they can be naturally combined, e.g.~one can do both sparsify the model and reduce arithmetic precision.
Both strategies reduce the memory footprint accordingly and are vital for the deployment of DNNs in many real-world applications. This is especially important as reduced memory requirements are one of the main contributing factors reducing the energy consumption \cite{Han2016,Chen2016,Horowitz2014}. Apart from that, model size reduction and sparsification techniques such as weight pruning~\cite{LeCun1989, Han2015, Han2016}, weight sharing~\cite{Ullrich2017}, knowledge distillation~\cite{Hinton2015} of special weight matrix structures~\cite{Iandola2016,Cheng2015,Yang2015} also impacts the computational demand measured in terms of number of arithmetic operations. Unfortunately, this reduction usually does not directly translate into savings of wall-clock time, as current hardware and software are not well-designed to exploit model sparseness \cite{Zhang2016}.
Instead, reducing parameter precision proves quite effective for improving execution time on CPUs \cite{Vanhoucke2011, 7471821} and specialized hardware such as FPGAs~\cite{Umuroglu2017}. When the precision of the inference process is driven to the extreme, i.e. assuming binary weights $\vm{W} \in \{-1,1\}$ or ternary weights $\vm{W} \in \{-1,0,1\}$ in conjunction with binary inputs and binary activation functions $\sigma(x) \in \{-1,1\}$, floating or fixed point multiplications are replaced by hardware-friendly logical XNOR and bitcount operations, i.e.\ DNNs essentially reduce to a logical circuit.
Training such discrete-valued DNNs\footnote{Due to finite precision, in fact any DNN is discrete valued. However, we use this term here to highlight the extremely low number of values.} is delicate as they cannot be directly optimized using gradient based methods. However, the obtained computational savings on today's computer architectures are of great interest, especially when it comes to human-machine interaction (HMI) systems where latency and energy efficiency plays an important role.\\
%In the sequel, we provide a literature overview of approaches that use reduced-precision computation to facilitate low-ressource training and/or testing.

%, as current hardware and software are not well-designed to exploit model sparseness \cite{Zhang2016}.
%Several efforts have been made to reduce the number of bits required to store the weights, and thereby dramatically reducing the computational %requirements and memory footprint of a NN \cite{Courbariaux:2014, Courbariaux:2015}. \emph{Binary Neural Networks} (BNN) use only a single bit per weight %\cite{Courbariaux:2016, Rastegari:2016, Hubara:2016, Zoehrer:apr18}.

Due to steadily decreasing cost of consumer electronics, many state-of-the-art HMI systems include \emph{multi-channel speech enhancement} (MCSE) as a pre-processing stage. In particular, beamformers (BF) spatially separate background noise from the desired speech signal. Common BF methods include the \emph{Minimum Variance Distortionless Response} (MVDR) beamformer \cite{Veen:apr88} and the \emph{Generalized Eigenvalue} (GEV) beamformer \cite{Warsitz:Jul07}. Both MVDR and GEV beamformers are frequently combined with DNN-based  \emph{mask estimators}, estimating a gain-mask to obtain the spatial \emph{Power Spectral Density} (PSD) matrices of the desired and interfering sound sources. Mask-based BFs are amongst state-of-the-art beamforming approaches \cite{Higuchi:Mar16, Heymann:mar16, Heymann:dec15, Erdogan:sep16, Pfei:dec19, chime4_submission, Pfei:mar17, Pfei:aug17, Zoehrer:apr18}. However, they are computational demanding and need to be trimmed down to facilitate the usage in low-resource applications.\\

In this paper, we investigate the trade-off between performance and resources in MCSE systems. 
We exploit Bi-directional Long Short-Term Memory (BLSTM) architectures to estimate a gain mask from noisy speech signals, and combine it with both the GEV and MVDR beamformers~\cite{Pfei:dec19}. We analyze the computational demands of the overall system and highlight techniques to reduce the computational load. In particular, we observe that the computational effort for mask estimation is by orders of magnitude larger compared to subsequent beamforming. Hence, we concentrate on efficient mask estimation in the remainder of the paper. We limit the numerical precision of the mask estimation DNN's weights and after each processing step in the forward pass (i.e. inference) to either 8, 4 or 1 bits. This makes the MCSE system both resource and memory efficient. We report both perceptual audio quality and speech intelligibility of the overall MCSE system in terms of SNR improvement and \emph{Word Error Rate} (WER) using the Google Speech-to-Text API \cite{pySpeechRecognition}. In particular, we use the WSJ0 corpus \cite{Paul1992} in conjunction with simulated room acoustics to obtain 6 channel data of multiple, moving speakers \cite{Habets2007}. When reducing the numerical precision in the forward pass, the system still yields competitive results for single speaker scenarios with a slightly decreased WER. Additionally, we show that reduced-precision DNNs can be readily exploited on today's hardware, by benchmarking the core operation of binary DNNs (BNNs), i.e.~binary matrix multiplication, on NVIDIA Tesla K80 and ARM Cortex-A57 architectures.

The paper is structured as follows. In Section \ref{sec:systemmodel} we introduce the MCSE system. We highlight both MVDR and GEV beamformers and introduce DNN-based mask estimators. Section \ref{sec:Efficiency} provides details about the computational complexity of the MCSE system. We introduce reduced-precision LSTMs and discuss efficient representations in detail.  In Section~\ref{sec:exp} we present experiments of the MCSE system. In particular, the experimental setup and the results in terms of SNR and WER accuracy are discussed. Section~\ref{sec:con} concludes the paper.

\section{Multi-Channel Speech Enhancement System}
\label{sec:systemmodel}

The acoustic environment of our MCSE system consists of up to $C$ independent sound sources, i.e. human speech or ambient noise. The sound sources may be non-stationary, due to moving speakers, and their spatial and temporal characteristics are unknown. 

\begin{figure}[!ht]
\centering
\includegraphics[width=0.5\textwidth]{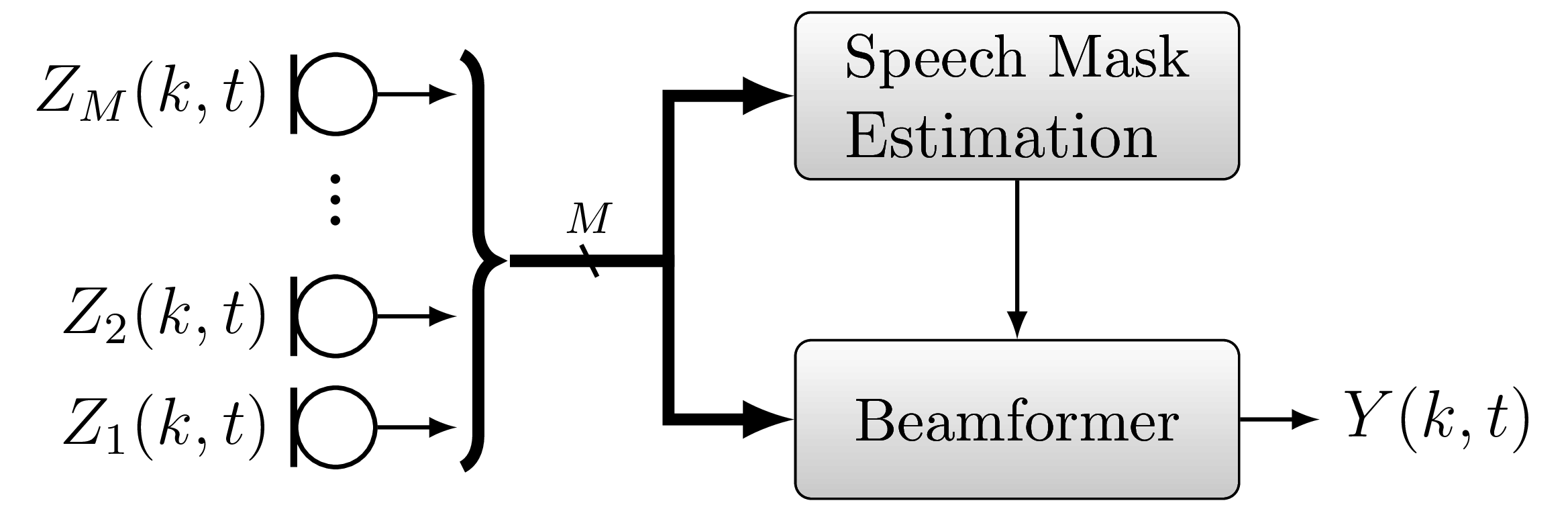}
\caption{System overview, showing the microphone signals $Z_m(k,t)$ and the beamformer output $Y(k,t)$ in frequency domain.}
\label{fig:system_overview}
\end{figure}

The speech enhancement system itself is composed of a circular microphone array with $M$ microphones, a DNN to estimate gain masks from the noisy microphone observations, and a \emph{broadband beamformer} to isolate the desired signal as shown in Figure (\ref{fig:system_overview}).

\noindent The signal at the microphones is a mixture of all $C$ sources, i.e. in \emph{short-time Fourier transform} (STFT) domain
\begin{equation}
  \vm{Z}(k,t) = \sum_{c=1}^{C} \vm{X}_c(k,t),
  \label{eq_mixture}
\end{equation}
\noindent where the time-frequency bins of all $M$ microphones are stacked into a $M \times 1$ vector $\vm{Z}(k,t) = \big[ Z_1(k,t), \ldots, Z_M(k,t) \big]^T$.  The vector $\vm{X}_c(k,t)$ represents the $c$\textsuperscript{th} sound source at all $M$ microphones at frequency bin $k = 1,\ldots,K$ and time frame $t$.\footnote{For the sake of brevity, the frequency and time frame indices will be omitted where the context is clear.} Each sound source is composed of a monaural recording $X_c(k,t)$ convolved with the \emph{Acoustic Transfer Function} (ATF) $\vm{A}_c(k,t)$, i.e.\

\begin{equation}
  \vm{X}_c(k,t) = \vm{A}_c(k,t) X_c(k,t),
  \label{eq_soundsource}
\end{equation}
\noindent where $\vm{A}_c(k,t)$ models the acoustic path from the $c$\textsuperscript{th} sound source to the  microphones, including all reverberations and reflections caused by the room acoustics \cite{roomacoustics1}. In the near field of the array, the ATFs can be modeled by a \emph{finite impulse response} (FIR) filter \cite{speechprocessing1}. The filter characteristics varies with the movement of the speaker, i.e. it is non-stationary. Without loss of generality, we specify the first source $c=1$ to be the desired source, i.e. $\vm{S}(k,t) = \vm{X}_1(k,t)$, and the interfering signal as the sum of the remaining sources, i.e. $\vm{N}(k,t) = \sum_{c=2}^{C} \vm{X}_c(k,t)$. The spatial PSD matrix for the desired signal is given as \cite{mimo1}

\begin{equation}
  \vm{\Phi}_{SS}(k) = \mathbb{E} \{ \vm{S}(k,t) \vm{S}^H(k,t) \},
  \label{eq_speech_psd}
\end{equation}
\noindent and for the interfering signal
\begin{equation}
  \vm{\Phi}_{NN}(k) = \mathbb{E} \{ \vm{N}(k,t) \vm{N}^H(k,t) \}.
  \label{eq_noise_psd}
\end{equation}

%\section{Beamforming}
%\label{sec:beamforming}
The aim of beamforming is to recover the desired source $\vm{S}(k,t)$ while suppressing the interfering sources $\vm{N}(k,t)$ at the same time. We use a \emph{filter and sum} beamformer \cite{micarray2}, where each microphone signal $Z_m(k,t)$ is weighted with the beamforming weights $W_m(k)$, prior to summation into the result $Y(k,t)$, i.e.
\begin{equation}
  Y(k,t) = \vm{W}^H(k) \vm{Z}(k,t),
\label{eq_bf_output}
\end{equation}
\noindent where $\vm{W}(k) = \big[ W_1(k), \ldots, W_M(k) \big]^T$.

\subsection{MVDR Beamformer}

The MVDR beamformer \cite{speechprocessing1, Shmulik:09} minimizes the signal energy at the output of the beamformer, while maintaining an undistorted response with respect to the \emph{steering vector} $\vm{v}_S(k)$, i.e. its weights $\vm{W}_{MVDR}$ are

\begin{equation}
  \vm{W}_{MVDR} = \frac{ \vm{\Phi}^{-1}_{NN} \vm{v}_S }{ \vm{v}_S^H \vm{\Phi}^{-1}_{NN} \vm{v}_S }.
\label{eq_mvdr}
\end{equation}

\noindent The steering vector $\vm{v}_S(k)$ guides the beamformer towards the direction of the desired signal. This direction can be determined using \emph{Direction Of Arrival} (DOA) estimation algorithms \cite{micarray1, Gannot:Aug01, Talmon:May09, Pfei:May14}. However, In real-world application this is sub-optimal, as it does not consider reverberations and multi-path propagations. Assuming that the PSD matrix of the desired source is known, the steering vector can be obtained in signal subspace \cite{Shmulik:09} using Eigenvalue decomposition (EVD) of the PSD matrix $\vm{\Phi}_{SS}(k)$. 
%\begin{equation}
%  \vm{\Phi}_{SS}(k) = \sum_{m=1}^M \vm{v}_{S_m} \vm{v}_{S_m}^H \lambda_{S_m},
%\label{eq_Phi_ss_evd}
%\end{equation}
%
%\noindent where $\lambda_{S_m}$ and $\vm{v}_{S_m}$ are the eigenvalues and eigenvectors of $\vm{\Phi}_{SS}$, respectively. 
In particular, the eigenvector belonging to the largest eigenvalue is used as steering vector $\vm{v}_S(k)$.

\subsection{GEV Beamformer}

An alternative to the MVDR beamformer is the GEV beamformer \cite{Warsitz:Jul07, Warsitz:08}. It determines the filter weights $\vm{W}$ to maximize the SNR $\xi$ at the beamformer output, i.e. 

\begin{equation}
  \vm{W}_{GEV} = \underset{\vm{W}}{\arg\max} \, \xi,
\label{eq_max_snr}
\end{equation}

\noindent where

\begin{equation}
  \xi = \frac{ \vm{W}^H \vm{\Phi}_{SS} \vm{W} }{ \vm{W}^H \vm{\Phi}_{NN} \vm{W} }.
\label{eq_max_snr2}
\end{equation}

\noindent  Eq. (\ref{eq_max_snr}) can be rewritten as a generalized Eigenvalue problem \cite{Warsitz:08}: \

\begin{equation}
\vm{\Phi}_{SS} \vm{W} = \xi \vm{\Phi}_{NN} \vm{W},
\label{eq_gev}
\end{equation}

\noindent where $\vm{W}$  is the eigenvector belonging to the largest eigenvalue of $ \vm{\Phi}_{NN}^{-1} \vm{\Phi}_{SS}$. To compensate for the amplitude distortions ~\cite{Warsitz:Jul07} of the beamforming filter $\vm{W}_{GEV}$, we choose the two reference implementations from ~\cite{Pfei:dec19}, i.e. the GEV-PAN, GEV-BAN postfilters.

%\noindent where $\zeta$ is an arbitrary complex scalar. Since, the response of the beamforming filter $\vm{W}_{GEV}$ shows distortions, i.e. $\vm{v}_S^H \vm{W}_{GEV} \neq 1$, the \emph{Blind Analytical Normalization} (BAN)~\cite{Warsitz:Jul07} or \emph{Phase Aware Normalization} (PAN)~\cite{Pfei:aug17} compensation factor have been proposed. The PAN factor turns the GEV beamformer into an MVDR beamformer. However, the GEV BF avoids the inversion of the noise PSD matrix $\vm{\Phi}_{NN}$ by using the Schur decomposition. This leads to improved numerical stability~\cite{Heymann:mar16}.

\subsection{PSD Matrix Estimation}
\label{sec_psd_est}
The spatial PSD matrix $\vm{\Phi}_{SS}(k)$ can be approximated using
\begin{equation}
  \vm{\hat{\Phi}}_{SS}(k,t) = \frac{1}{L} \sum\limits_{l=t-L/2}^{t+L/2} \vm{Z}(k,l) \vm{Z}^H(k,l) p(k,l),
\label{eq_Sss_approx}
\end{equation}
and the gain mask $p(k,t)$ for the speech signal. Analogously, $\vm{\hat{\Phi}}_{NN}(k,t)$ can be estimated using the gain mask $q(k,t)$ for the interfering signal. Note that the window length $L$ defines the number of time frames used for estimating the PSD matrices. For moving sources, $L$ has to be sufficiently large to obtain well estimated PSD matrices. If $L$ is too large, the estimated PSD matrices might fail to adapt quickly enough to changes in the spatial characteristics of the moving sources. An alternative is provided by recursive estimation, i.e.\

\begin{equation}
\begin{aligned}
  \vm{\hat{\Phi}}_{SS}(k,t) &\leftarrow \vm{\hat{\Phi}}_{SS}(k,t-1) \big[1 - p(k,t) \big] \\
                            &\phantom{\leftarrow} + \vm{Z}(k,t) \vm{Z}^H(k,t) p(k,t).
\end{aligned}
\label{eq_Sss_approx2}
\end{equation}

\noindent This \emph{online processing} \cite{Bodekker:apr2018} allows to adapt the MVDR or GEV beamformer at each time frame $t$. This recursive estimation is initialized using Eq. (\ref{eq_Sss_approx}).

\subsection{Recursive Eigenvector Tracking}

If Eq. (\ref{eq_Sss_approx}) is used, the generalized Eigenvalue decomposition in Eq. (\ref{eq_gev}) has to be performed for every time-frequency bin. This expensive operation can be circumvented by \emph{recursive Eigenvector tracking} using Oja's method \cite{haykin:2009neural}; i.e.\

\begin{equation}
\begin{aligned}
  \vm{W}'(k,t) & \leftarrow \vm{W}'(k,t-1) \\
  & \phantom{\leftarrow} - \alpha \vm{\hat{\Phi}}_{NN}(k,t)  \vm{W}'(k,t-1) \\
  & \phantom{\leftarrow} + \alpha \vm{\hat{\Phi}}_{SS}(k,t)  \vm{W}(k,t-1)\quad\mbox{, and} \\
  \vm{W}(k,t) & \leftarrow \frac{ \vm{W}'(k,t) }{ || \vm{W}'(k,t) ||_2 }.
\end{aligned}
\label{eq_oja}
\end{equation}

Note that the GEV-PAN and GEV-BAN postfilters from \cite{Pfei:dec19} are not affected by the normalization operation in Eq. (\ref{eq_oja}). When using the MVDR beamformer, tracking the largest eigenvector of $\vm{\hat{\Phi}}_{SS}(k,t)$ is done in a similar fashion.

\subsection{DNN-based Speech Mask Estimation}
\label{dnn:architecture}

The DNN used to estimate the gain mask for the beamformer uses the noisy microphone observations $\vm{Z}(k,t)$ as features. In particular, the features per time-frequency-bin are defined as $x(t,k) = [\Re{\vm{\bar{Z}}(k,t)}, \Im{\vm{\bar{Z}}(k,t)}]$, where $\vm{\bar{Z}}(k,t)$ is a whitened and phase-normalized version of $\vm{Z}(k,t)$. Further details on whitening can be found in \cite{Pfei:dec19}. For $M$ microphones, $x(t,k)$ contains $2M$ real-valued elements. The DNN processes $K$ frequency bins at a time, hence each time frame $t$ uses the feature vector $\vm{x}(t) = [x(t,0), x(t,1), \ldots, x(t,K-1)]$ as input. It contains $2MK$ elements.

\begin{figure}[!htb]
\centering
\includegraphics[width=0.5\textwidth]{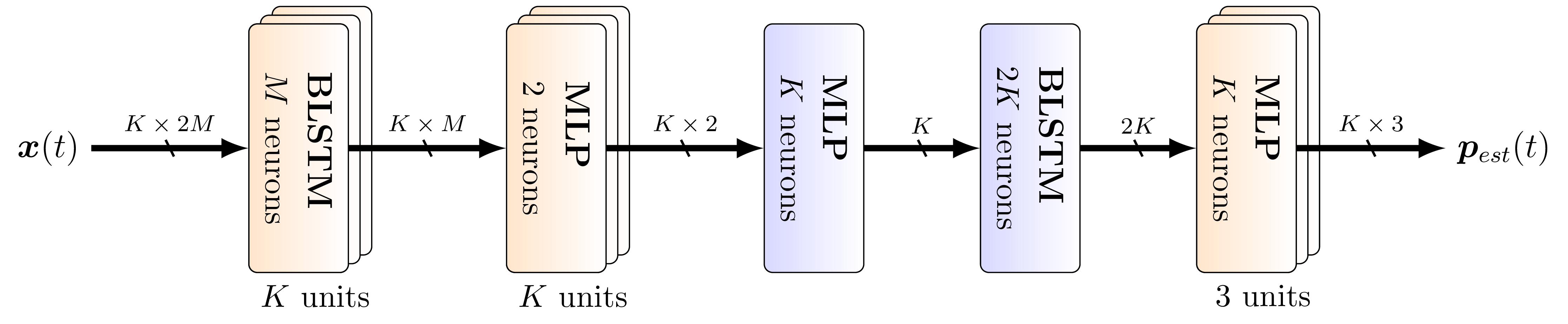}
\caption{Mask estimation DNN.}
\label{fig:evs_architecture}
\end{figure}

Figure \ref{fig:evs_architecture} shows the architecture of the DNN consisting of \emph{Dense layers} and BLSTM units. Similar architectures for speech mask estimation can be found in \cite{Deng:2010, Heymann:mar16, Zoehrer:Aug15, Pfei:mar17}.

The first BLSTM layer consists of two separate LSTM units~\cite{hochreiter1997long}, each with $2M$ neurons for each frequency bin. While the first LSTM processes the data in forward direction (i.e. one time frame after another), the second LSTM operates in backward direction. The output of both LSTMs is then concatenated to an intermediate vector with $2K$ elements.  The second and third layer consists of a Dense layer. The first three layers reduce the feature vector size from $2M$ elements per time-frequency bin down to 1. Note that those layers have very few weights, as they consist of $K$ independent units with $M$ neurons each. The fourth layer is a BLSTM processing all $K$ frequency bins at a time. Finally, three separate Dense layers are used.
The first dense layer estimates the mask for the desired source, the second estimates the mask for the interfering sources, and the third estimates the mask for time-frequency bins which are not assigned to the other two classes. The activation function of this layer is a softmax, so that the sum of each of the three masks is 1 for each time-frequency bin, i.e. $\sum_{i=1}^3 p_{est}(k,t,i) \overset{!}{=} 1$.

\section{Computational Efficiency of the MCSE System}
\label{sec:Efficiency}

\subsection{Complexity analysis of MCSE system}\label{sec:computational_savings}

Table \ref{tab:eff_maskestimator} shows both the computational complexity and the number of \emph{multiply-and-accumulate} (MAC)  operations for the proposed DNN-based mask estimator (cf. Section 2). Overall, 5562e6 MAC operations are needed to compute a gain-mask given a multi-channel signal with $M=6$ microphones, $K=513$ frequency bins and $T=500$ frames. Table \ref{tab:eff_bf} shows the MAC operations of a static and dynamic beamformer, needed to infer the target speech. Static beamformers, which do not track moving targets have a reduced computational overhead compared to dynamic variants, computing the beamforming weight for every time-step. However, the overall computational complexity is orders of magnitude lower compared to the DNN-based mask estimator. This indicates that significant computational savings can be obtained when optimizing DNNs with respect to resource efficiency.

\begin{table}[htb]
\caption{Computational complexity for the DNN-based mask estimator.}
  \centering
  \scriptsize
\begin{tabular}{|l|l|r|r|}
\hline
Layer & Shape & Weights & MAC \\
\hline
BLSTM & $16 \times K \times 2M \times M$ & 590976 & 295e6\\
Dense layer & $K \times M \times 2$ & 6156 & 3e6\\
Dense layer & $2K \times K$ & 526338 & 263e6 \\
BLSTM & $16 \times K \times 2K$ & 8421408 & 4211e6\\
Dense layer & $3 \times 2K \times K$ & 1579014 & 790e6\\
\hline
Total & & 11123892 & 5562e6\\
\hline
\end{tabular}

\label{tab:eff_maskestimator}
\end{table}

\begin{table}[htb]
\caption{Computational complexity of a static- and dynamic GEV beamformer.}
  \centering
  \scriptsize
\begin{tabular}{|l|l|l|r|}
\hline
Mode & Layer & Complexity & MAC \\
\hline
static & Eq. 10 & $\mathcal{O}(T*2*K*M^2)$ & 18e6 \\
static & Eq. 9 & $\mathcal{O}(K*M^3)$ & 0.1e6 \\
\hline
Total & & & 18.1e6 \\
\hline
\hline
dynamic & Eq. 10 & $\mathcal{O}(T*2*K*M^2)$ & 18e6 \\
dynamic & Eq. 9  & $\mathcal{O}(T*K*M^3)$ & 55e6 \\
\hline
Total & & & 73e6 \\
\hline
\end{tabular}
\label{tab:eff_bf}
\end{table}

Reducing the precision of the DNN-based mask estimators reduces the computational complexity and memory consumption of the overall MCSE system. Reduced precision DNNs can be realized via bit-packing\footnote{https://github.com/google/gemmlowp} schemes, with the help of processor specific GEMM instructions \cite{Abdelfattah:2019} or can be implemented on a DSP or FPGA. 

Computational savings for various 8-bit DNN models on both ARM processors and GPUs have been reported in \cite{Jahre:2017, Umuroglu:2017, schindler:2018, Abdelfattah:2019}. In particular, \cite{Vanhoucke2011} reported that speech recognition performance is maintained when quantizing the neural network parameters to 8 bit fixed-point, while the  system runs 3 times faster on a x86 architecture.  

In order to demonstrate the advantages that binary computations achieve on other general-purpose processors, we implemented matrix-multiplication operators for NVIDIA GPUs and ARM CPUs. BNNs can be implemented very efficiently as 1-bit scalar products, i.e.~multiplications of two vectors $\mathbf{x}$ and $\mathbf{y}$ of length $N$ reduce to bit-wise \textit{xnor()} operation, followed by counting the number of set bits with \textit{popc()}, i.e.\
\begin{equation}\label{eq:xnor}
\mathbf{x}\cdot \mathbf{y} = N-2*popc(xnor(\mathbf{x},\mathbf{y})), x_i, y_i \in\{-1, +1\}\quad\forall{i},
\end{equation}
where $x_i$ and $y_i$ denote the $i\textsuperscript{th}$ element of $\mathbf{x}$ and $\mathbf{y}$, respectively.
We use the matrix-multiplication algorithms of the MAGMA and Eigen libraries and replace float multiplications by \textit{xnor()} operations, as depicted in Equation~\eqref{eq:xnor}. 
Our CPU implementation uses NEON vectorization in order to fully exploit SIMD instructions on ARM processors. 
We report execution time of GPUs and ARM CPUs in Table \ref{tab:performance_metrics}.  
As can be seen, binary arithmetic offers considerable speed-ups over single-precision with manageable implementation effort.  This also affects energy consumption since binary values require less off-chip accesses and operations. Performance results of x86 architectures are not reported because neither SSE nor AVX ISA extensions support vectorized \textit{popc()}. 

\begin{table}[htb]
	\caption{Performance metrics for matrix $\cdot$ matrix multiplications on a NVIDIA Tesla K80 and ARM Cortex-A57.}
  \centering
  \scriptsize
  \begin{tabular}{|l|l l l l|}
      \hline
      arch  & matrix size & time (float32) & time (binary) & \bf speed-up\\
      \hline
	  GPU & 256 & 0.14ms & 0.05ms & \bf2.8\\	
	  GPU & 513 & 0.34ms & 0.06ms & \bf5.7\\
	  GPU & 1024 & 1.71ms & 0.16ms & \bf10.7\\	
	  GPU & 2048 & 12.87ms & 1.01ms & \bf12.7\\	
      \hline
	  ARM & 256 & 3.65ms & 0.42ms & \bf8.7\\	
	  ARM & 513 & 16.73ms & 1.43ms & \bf11.7\\
	  ARM & 1024 & 108.94ms & 8.13ms & \bf13.4\\
	  ARM & 2048 & 771.33ms & 58.81ms & \bf13.1\\
      \hline	
    \end{tabular}

    \label{tab:performance_metrics} 

\end{table}

\subsection{Reduced Precision DNNs}
We exploit reduced-precision weights and limit the numerical precision of a DNN-based mask estimator to either 8- or 4 bit fixed-point representations or to binary weights. Recently, there has been numerous extensions to train DNNs with limited precision ~\cite{dorefa, wage, NIPS2018,Courbariaux2015a}.

\subsubsection{DNN with Low-precision Weights}
The weights and activations of a DNN often lie within a small range, making it possible to introduce quantization schemes. Implementations like \cite{OttLZLB16,7471821} use reduced precision for their DNN's weights. In \cite{Vanhoucke2011}, an improvement of inference speed of factor 3 for fixed-point implementation on a general purpose hardware has been reported. Hence, we consider a fixed-point representation of the computed values in the \emph{forward pass} of our DNN~\cite{Courbariaux2015a}. In particular, we use 8- and 4 bit weights, which represent the Q2.6 and Q2.2 fractional formats, respectively. After each layer, we use batch normalization to ensure the activations to fit within $\in{[-2, +2]}$. The accumulation of the values in the dot products and the batch normalization are performed with high precision, while the multiplication is performed at lower precision. 

\begin{figure}[!htb]
\centering
\includegraphics[width=0.4\linewidth]{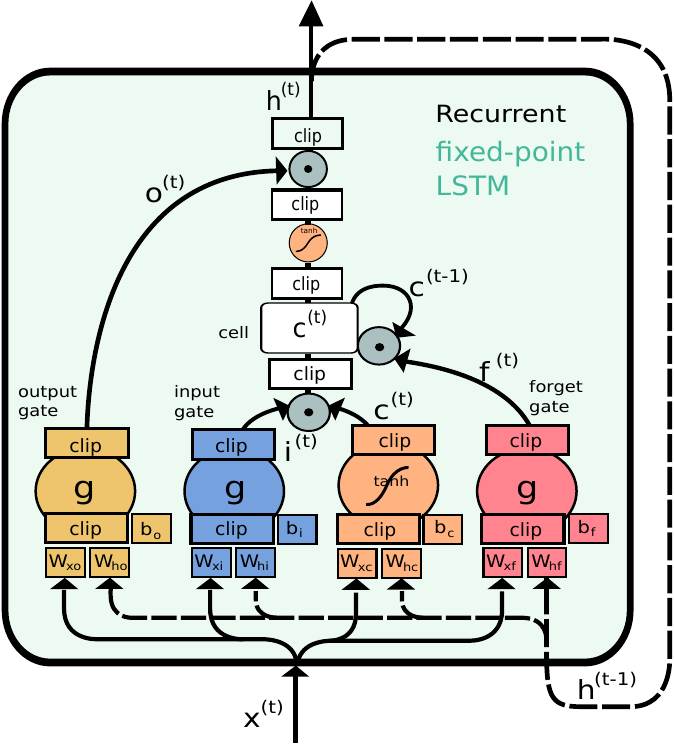}
\caption{Reduced precision LSTM cell, using bit-clipping to either 4- or 8 bit fixed-point representation after each mathematical operation.}
\label{fig:fixed_point_lstm}
\end{figure}

During training we compute the gradient and update the weights using float32, while the precision is only reduced accordingly in the forward pass\footnote{The derivative is computed with respect to the quantized weights as in~\cite{Courbariaux2015b,dorefa,OttLZLB16}.}. This is known as \emph{straight through estimator} (STE)~\cite{Courbariaux2015b,dorefa}, where the parameter update is performed in full-precision. Usually, when deploying the DNN in an application, only the forward-pass calculations are required. Hence, the reduced-precision weights can be used, reducing memory requirements by a factor of 4 or 8 compared to 32-bit weight representations. Figure \ref{fig:fixed_point_lstm} shows a reduced-precision LSTM cell. Besides the well-known gating and \textit{vector}$\cdot$\textit{matrix} computations of LSTMs, bit clipping operations are introduced after each mathematical operation. Details of the LSTM cell can be found in~\cite{hochreiter1997long}.

%\clearpage

\subsubsection{DNN with Binary Weights}
In \cite{Courbariaux2015b}, binary-weight DNNs are trained using the STE, i.e. deterministic and stochastic rounding is used during forward propagation, and the full-precision weights are updated based on the gradients of the quantized weights. In \cite{Hubara2016}, STE is used to quantize both the weights and the activations to a single bit and sign functions respectively.  
~\cite{Li2016} trained ternary weights $\vm{W} \in \{-a,0,a\}$ by setting weights below or above a certain threshold $\Delta$ to $\pm a$, or zero otherwise. %This has been extended in %\cite{Zhu2017} to $\vm{W} \in \{-a,0,b\}$, where the thresholds $a>0$ and $b>0$ are trained %using gradient updates.
This has been extended in \cite{Zhu2017} to ternary weights $w \in \{-a,0,b\}$ by learning the factors $a>0$ and $b>0$ using gradient updates and a different threshold $\Delta$ has been applied.

%This further reduces the computational burden as floating-point multiplications and additions are reduced to hardware-friendly logical XNOR and bitcount operations.

\begin{figure}[!htb]
\centering
\includegraphics[width=0.4\linewidth]{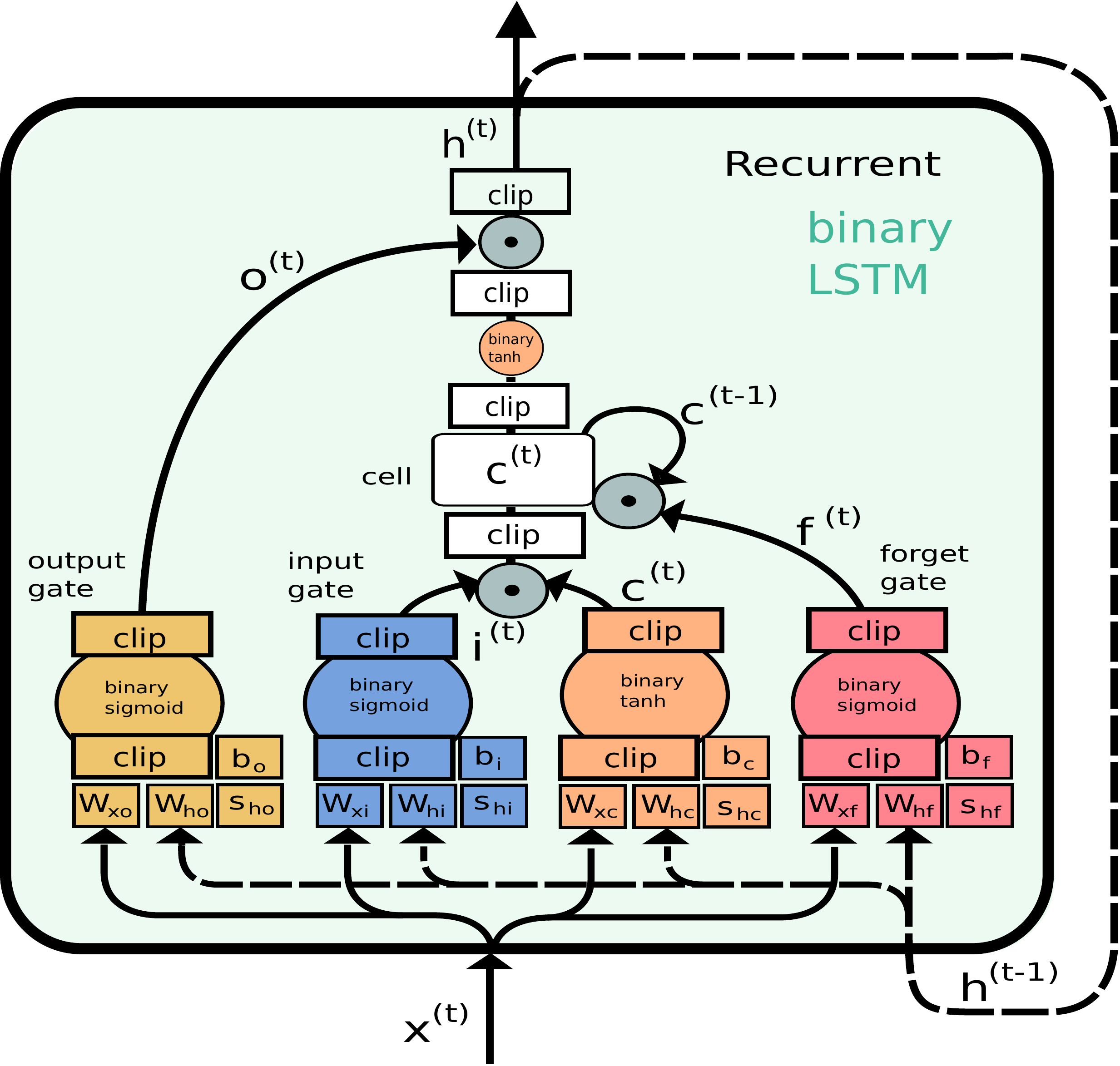}
\caption{Binary LSTM, using both binary weights and binary activation functions and scaling parameters $\{ \Sih, \Soh, \Sch, \Sfh \}$.}
\label{fig:binary_mlp}
\end{figure}

When dealing with recurrent architectures such as LSTMs, ~\cite{OttLZLB16} observed that recent reduced-precision techniques for BNNs~\cite{Courbariaux2015b,Hubara2016} cannot be directly extended to recurrent layers. In particular, a simple reduction of the precision of the forward pass to 1 bit in the LSTM layers suffers from severe performance degradation as well as the vanishing gradient problem. In \cite{Zhu2017, Ardakani:2019} batch-normalization and a weight-scaling is applied to the recurrent weights to overcome this problem. We adopt this approach, i.e. introducing a trainable scaling parameter $s$, which maps the range of the recurrent activations to $\{-s,s\}$. Hence, each of the recurrent weight matrices $\vm{W}_{ho}, \vm{W}_{hi}, \vm{W}_{hc}$ and $\vm{W}_{hf}$ has its own scaling factor, i.e. $\{\Sih, \Soh, \Sch, \Sfh \}$. See also Fig. \ref{fig:binary_mlp}. This limits the recurrent weights to small values, preventing the LSTM to reach unstable states, i.e.\ avoids accumulating the cell states to large numbers. For binary weights, the LSTM cell equations are given as:

\begin{subequations}
  \begin{align}
    \igate &= \sigma_b(\Wix \xt + (\Wih \hidtLast) \odot \Sih + \bi)
    \label{eq:lstm:i}\\
    \fgate &= \sigma_b(\Wfx \xt + (\Wfh \hidtLast) \odot \Sfh + \bfgate)
    \label{eq:lstm:f}\\
    \ogate &= \sigma_b(\Wox \xt + (\Woh \hidtLast) \odot \Soh + \bo)
    \label{eq:lstm:o}   
  \end{align}
  \begin{align}
  \cellInt &= \tanh_b(\Wcx \xt + (\Wch \hidtLast) \odot \Sch + \bcin)
    \label{eq:lstm:c_in}\\
    \cellt &= \fgate \odot \celltLast + \igate \odot \cellInt
    \label{eq:lstm:c}\\
    \hidt &= \ogate \odot \tanh_b(\cellt),
  \label{eq:lstm:h}
  \end{align}
\end{subequations}
where $\sigma_b(\cdot)$ and $\tanh_b(\cdot)$ are a binary version (i.e. hard sigmoid and sign function~\cite{Courbariaux2015b}) of the well-known \textit{sigmoid} and \textit{tanh} activation functions. The weights $\vm{W}$ and biases $\vm{b}$ are the binary network parameters (i.e.\ with values of $\{-1,1\}$), and $\{\Sih, \Soh, \Sch, \Sfh \}$ are the scaling parameters for the recurrent network weights.

\section{Experiments}
\label{sec:exp}

\subsection{Experimental Setup}
\begin{figure} [htb!]
  \centering
  \includegraphics[width=0.5\textwidth]{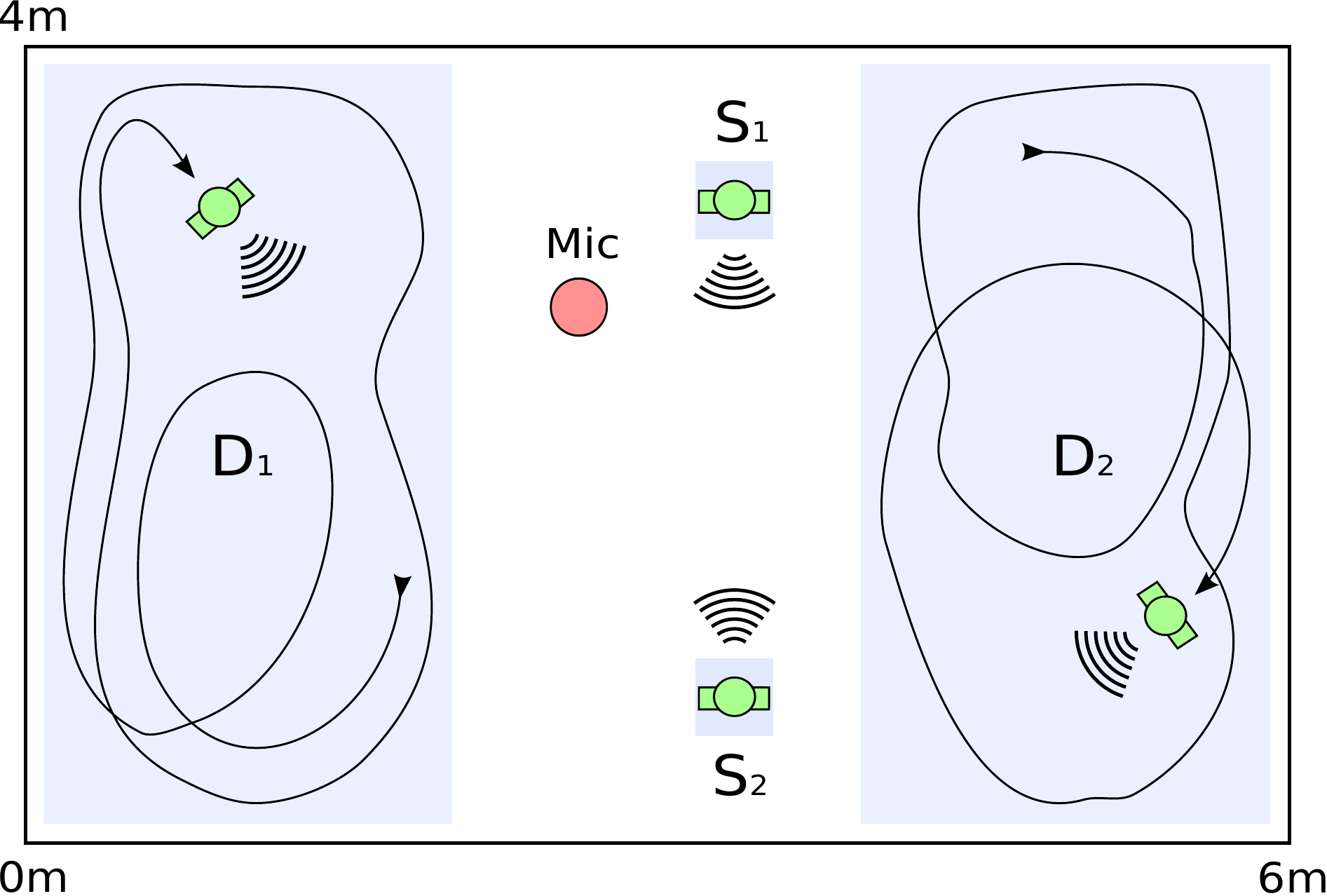}     
  \caption{Shoebox model of a living room showing two stationary sound sources \ssf{S}\textsubscript{1} and \ssf{S}\textsubscript{2}, and two dynamic sound sources \ssf{D}\textsubscript{1} and \ssf{D}\textsubscript{2}. The microphone array  (\ssf{Mic}) is visualized as red circle.}
  \label{fig:shoebox}
\end{figure}

The performance of the multi-channel speech enhancement system is demonstrated by simulating a typical living room scenario with two static speakers \ssf{S}\textsubscript{1} and \ssf{S}\textsubscript{2}, two moving speakers \ssf{D}\textsubscript{1} and \ssf{D}\textsubscript{2}, and an isotropic background noise source \ssf{I} similar as in \cite{Pfei:dec19}. The floor plan of the setup is shown in Figure \ref{fig:shoebox}. The circular microphone array with $M=6$ microphones and a diameter of $86mm$ is shown in red labeled as \emph{Mic}. Head movements of the static speakers \ssf{S}\textsubscript{1} and \ssf{S}\textsubscript{2} are simulated by random 3D position changes within $20cm$. The trajectory of the moving speakers \ssf{D}\textsubscript{1} and \ssf{D}\textsubscript{2} random within a region of 2m $\times$ 4m on both sides of the microphone array. The movement velocity is constant at $0.5\frac{m}{s}$. 

\noindent We specify five scenarios for our experiments using this shoebox model:

\begin{enumerate}
\item Random vs. isotropic (\fsf{R}-\fsf{I}): A static speaker with head movements is the \emph{random} source. The position is randomly selected in the room for each new utterance to prevent the model from learning the position of the speaker.
\item Static1 vs. isotropic (\fsf{S}\textsubscript{1}-\fsf{I}): A stationary speaker at fixed position \ssf{S}\textsubscript{1} and an isotropic background noise are used in this scenario. The head movements cause a varying phase especially at higher frequencies.
\item Static1 vs. static2 + isotropic (\fsf{S}\textsubscript{1}-\fsf{S}\textsubscript{2}\fsf{I}): Two simultaneously talking speakers at position \ssf{S}\textsubscript{1} and \ssf{S}\textsubscript{2} embedded in isotropic background noise are used in this scenario. 
\item Dynamic1 vs. isotropic (\fsf{D}\textsubscript{1}-\fsf{I}): The speaker moving in region \fsf{D}\textsubscript{1}  has to be tracked in the presence of ambient background noise. This challenges the tracking capabilities of the DNN mask estimation.
\item Dynamic1 vs. dynamic2 + isotropic (\fsf{D}\textsubscript{1}-\fsf{D}\textsubscript{2}\fsf{I}): The separation capabilities of two speakers moving in \fsf{D}\textsubscript{1} and \fsf{D}\textsubscript{2} embedded in background noise is analysed.
\end{enumerate}
These experimental setups are summarized in  Table \ref{tab:experiments}:\\

\begin{table}[htb!]
  \caption{Experimental setups using a virtual shoebox model.}
\centering
  \begin{footnotesize}
  \begin{tabular}{|c|c|c|}
  \hline
  Experiment \# & Desired source & Interfering source(s) \\
  \hline
  1 & random \fsf{R} & isotropic \fsf{I} \\
  2 & \fsf{S}\textsubscript{1} & isotropic \fsf{I}\\
  3 & \fsf{S}\textsubscript{1} & \fsf{S}\textsubscript{2}, isotropic \fsf{I} \\  
  4 & \fsf{D}\textsubscript{1} & isotropic \fsf{I}\\
  5 & \fsf{D}\textsubscript{1} & \fsf{D}\textsubscript{2}, isotropic \fsf{I}\\
  \hline
  \end{tabular}
  \label{tab:experiments}
  \end{footnotesize}
\end{table}

\subsection{Data Generation}

We use the \emph{Image Source Method} (ISM) \cite{Habets2007, pyroomacoustics} to simulate the ATFs in Eq. (\ref{eq_soundsource}). This enables to generate multi-channel recordings from a monaural source. The room is modeled as shoebox with a reflection coefficient of $\beta = 0.85$ for each wall. The reflection order is $10$ which results in a reverberation time of $~500ms$. We generate a new set of ATFs every $32ms$  for the moving sources. The isotropic background noise is determined as
\begin{equation}
  \vm{X}_n(k,t) = \vm{U}(k,t) X_n(k,t),
  \label{eq:isotropic}
\end{equation}

\noindent where $X_n(k,t)$ is the monaural noise source, $\vm{U}(k,t) = \vm{E}(k) \vm{\Lambda}(k)^{0.5} \phantom{.} \exp{i \vm{\varphi}(k,t)}$, $\vm{\Lambda}(k)$ and $\vm{E}(k)$ denotes the eigenvalue and eigenvector matrices of the spatial coherence matrix $\vm{\Gamma}(k)$ for a spherical sound field \cite{roomacoustics1}. The $M \times 1$ vector $\vm{\varphi}(k,t)$ denotes a uniformly distributed phase between $-\pi, \dots, \pi$. 

\subsection{Training and Testing}

We use 12776 utterances from the \textit{si\_tr\_s} set of the WSJ0 \cite{Paul1992} corpus for the speech sources in Eq. (\ref{eq_soundsource}) for training. Additionally, 20 hours of different sound categories from YouTube \cite{pyTube} are used as isotropic background noise. All recordings are sampled at 16kHz and converted to the frequency domain with  $K=513$ bins and 75\% overlapping blocks. The sources are mixed with equal volume. For testing, we use 2907 utterances from the \textit{si\_et\_05} set of the WSJ0 corpus mixed with Youtube noise. 

The ground truth gain masks required for training can be obtained for the desired signal as:
\begin{equation}
  p_{opt}(k,l,1) = ||\vm{S}(k,t)||_2 > \max \big( ||\vm{N}(k,t)||_2 , \epsilon(k) \big).
\end{equation}

\noindent The mask for the interfering signals is given as:
\begin{equation}
  p_{opt}(k,l,2) = ||\vm{N}(k,t)||_2 > \max \big( ||\vm{S}(k,t)||_2 , \epsilon(k) \big).
\end{equation}

\noindent The weak signal components, which do not contribute to any of the PSD matrices, are obtained as:
\begin{equation}
  p_{opt}(k,l,3) = 1 - p(k,l,1) - p(k,l,2).
\end{equation}

\noindent Parameter $\epsilon(k)$ specifies the amount of energy per frequency bin $k$ required for the signal to be assigned to either the desired or interfering class label. Note that the calculation of the ground truth masks requires the corresponding signal energies $||\vm{S}(k,t)||_2$ and $||\vm{N}(k,t)||_2$ to be known, which is why we used the ISM rather than existing multi-channel speech databases such as \cite{Lincoln:nov05}. 

By setting $\sum_{i=1}^3 p_{est}(k,t,i) = 1$ for each time-frequency bin, we can use the \emph{cross-entropy} as loss function. For each (B)LSTM or dense layer a \textit{tanh} activation and batch normalization~\cite{Ioffe:2015} is applied.
We train for each of the five scenarios in Table \ref{tab:experiments} a separate DNN. Model optimization is done using stochastic gradient descent with ADAM \cite{Kingma:Jul15} using the cross-entropy between the optimal binary mask $p_{opt}$ and the estimated mask $p_{est}$ of the respective model. To avoid overfitting, we use early stopping by observing the error on the validation set every 20 epochs.

\subsection{Performance evaluation}

We use three different beamformers: the \ssf{MVDR}, \ssf{GEV-BAN} and \ssf{GEV-PAN} (see Section \ref{sec:systemmodel}) for each gain mask. The estimates of the PSD matrices are obtained using Eq. (\ref{eq_Sss_approx}), where $L = 32$ blocks. We apply the \emph{BeamformIt} toolkit \cite{Anguera2007} as baseline. It uses DOA estimation \cite{micarray1} followed by a MVDR beamformer. To evaluate the performance of the enhanced signals $Y(k,t)$, we use the Google Speech-to-Text API \cite{pySpeechRecognition} to perform \emph{Automatic Speech Recognition} (ASR). Furthermore, we determine the SNR improvement as: 

\begin{equation}
  \Delta SNR = 10log_{10} \frac{ \sum_{K,L} |Y(k,t)p_{opt}(k,l,1)|^2 }{ \sum_{K,L} |Y(k,t)p_{opt}(k,l,2)|^2 } \phantom{=} - 10log_{10} \frac{ \sum_{K,L} ||\vm{Z}(k,t)p_{opt}(k,l,1)||^2_2 }{ \sum_{K,L} ||\vm{Z}(k,t)p_{opt}(k,l,2)||^2_2 },
\label{eq_delta_snr}
\end{equation}

\noindent where the optimal \emph{binary} mask $p_{opt}$ is used to measure the energy of the desired and interfering components in the beamformer output $Y(k,t)$ and the noisy inputs $\vm{Z}(k,t)$, respectively. The $\Delta SNR$ can be computed without having access to the beamforming weights $\vm{W}(k,t)$, as is the case of the BeamformIt toolkit.

\subsection{Results}

While improvements of memory footprint and computation time are independent of the underlying tasks, the prediction accuracy highly depends on the complexity of the data and the used neural network. Simple data sets allow for aggressive quantization without affecting prediction performance significantly, while binary quantization results in severe prediction degradation on more complex data sets.\\

\begin{figure} [htb!]
\centering
\begin{tabular}{lr}

(a) & \makecell{\includegraphics[width=0.4\linewidth, trim={50px 10px 100px 30px},clip]{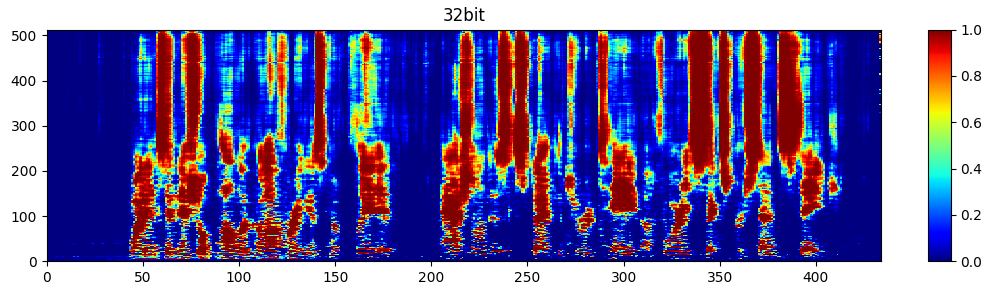}} \\

(b) & \makecell{\includegraphics[width=0.4\linewidth, trim={50px 10px 100px 30px},clip]{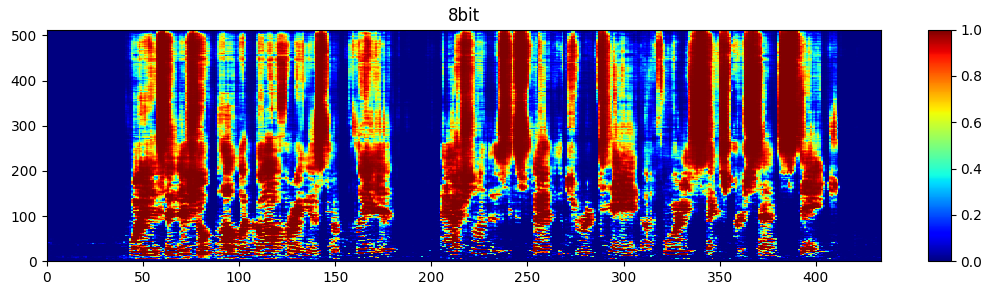}} \\

(c) & \makecell{\includegraphics[width=0.4\linewidth, trim={50px 10px 100px 30px},clip]{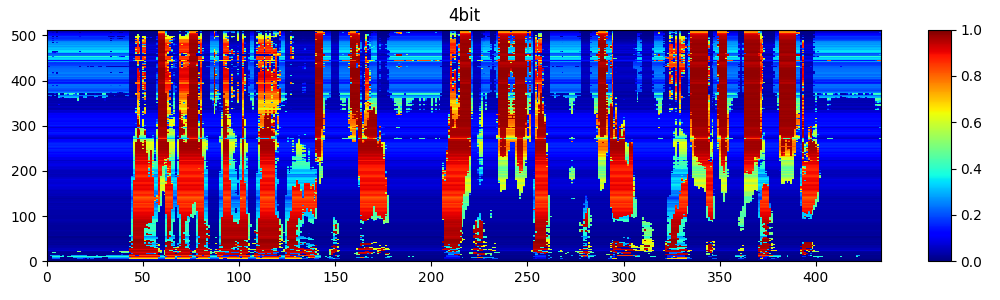}} \\

(d) & \makecell{\includegraphics[width=0.4\linewidth, trim={50px 10px 100px 30px},clip]{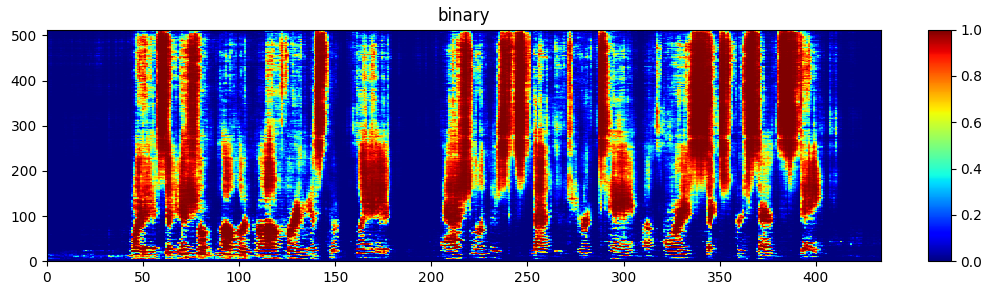}} \\

\end{tabular}

  \caption{Speech mask estimation $p_{est}(k,l,1)$ using (a) 32-, (b) 8-, (c) 4-, and (d) 1-bit DNNs.}
  \label{fig:mask_plot}
\end{figure}

Figure \ref{fig:mask_plot} shows speech mask estimations $p_{est}(k,l,1)$ using (a) 32-, (b) 8- (c) 4- and (d) 1-bit DNNs from the mixture of scenario (\fsf{S}\textsubscript{1}-\fsf{I}) of the WSJ0 utterance \textit{``When its initial public offering is completed Ashland is expected to retain a 46\% stake''} from \textit{si\_et\_05}. As noted in Section~\ref{dnn:architecture}, the activation function of the output layer is a full-precision softmax function. The reduction of the weight precision introduces artifacts in (b), (c) and (d).

\begin{figure} [htb!]
\centering
\begin{tabular}{lr}

(a) & \makecell{\includegraphics[width=0.4\linewidth, trim={50px 20px 100px 30px},clip]{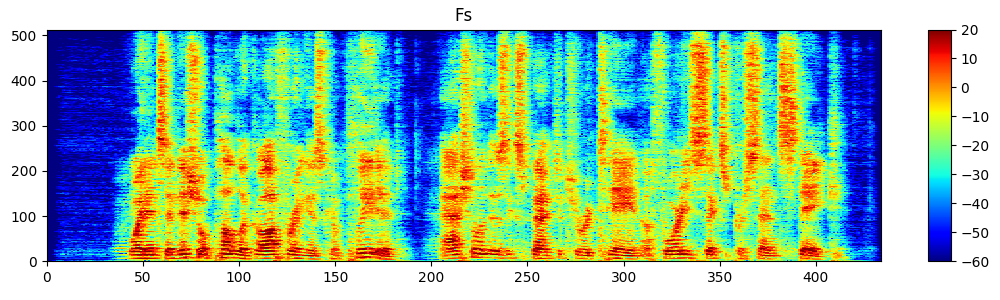}} \\

(b) & \makecell{\includegraphics[width=0.4\linewidth, trim={50px 10px 100px 30px},clip]{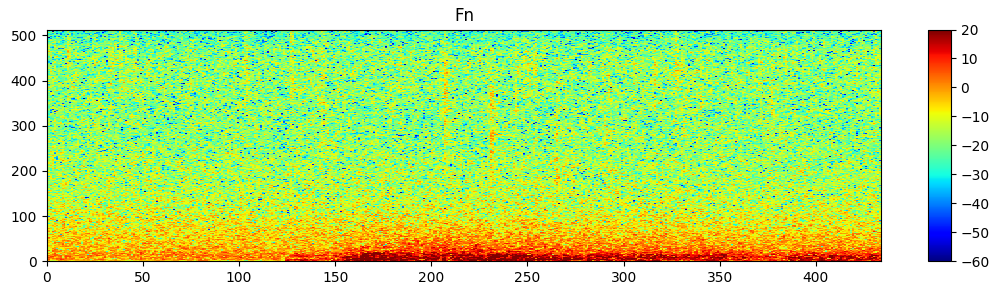}} \\

(c) & \makecell{\includegraphics[width=0.4\linewidth, trim={50px 10px 100px 30px},clip]{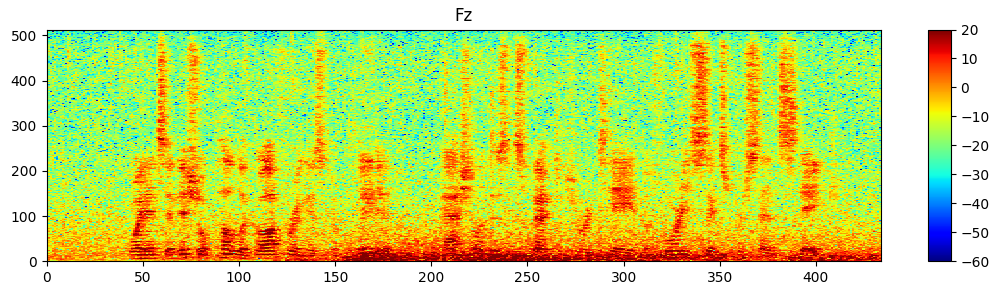}} \\

(d) & \makecell{\includegraphics[width=0.4\linewidth, trim={50px 10px 100px 30px},clip]{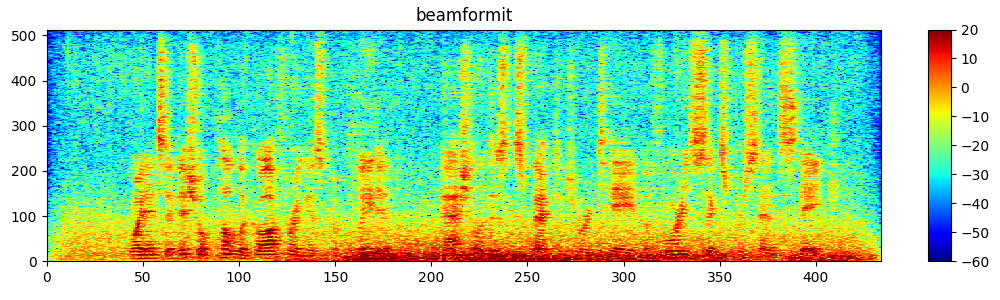}} \\

(e) & \makecell{\includegraphics[width=0.4\linewidth, trim={50px 10px 100px 30px},clip]{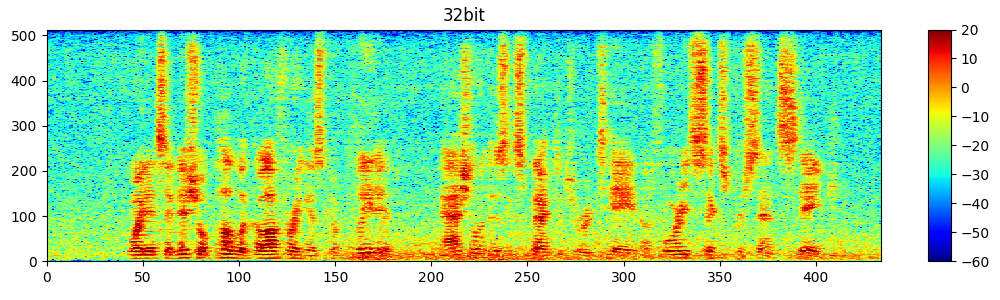}} \\

(f) & \makecell{\includegraphics[width=0.4\linewidth, trim={50px 10px 100px 30px},clip]{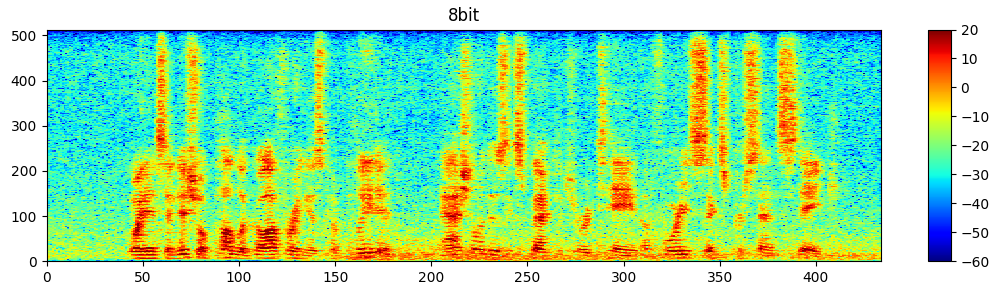}} \\

(g) & \makecell{\includegraphics[width=0.4\linewidth, trim={50px 10px 100px 30px},clip]{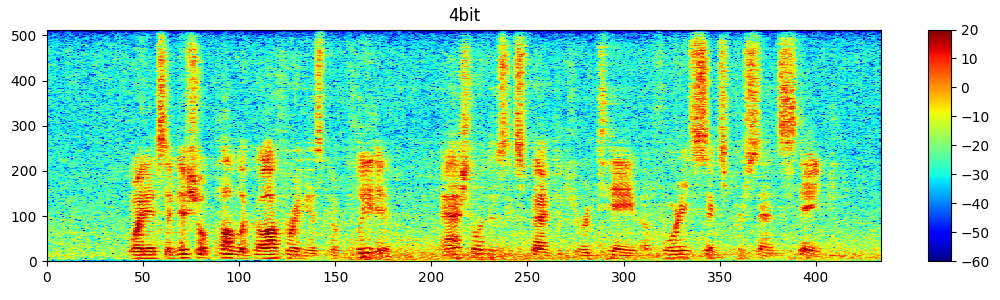}} \\

(h) & \makecell{\includegraphics[width=0.4\linewidth, trim={50px 10px 100px 30px},clip]{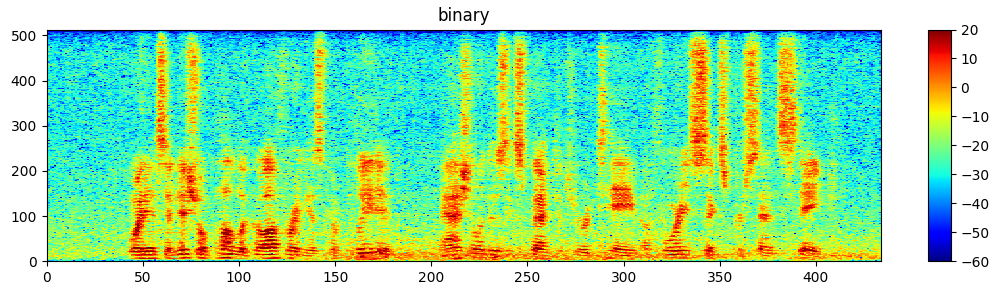}} \\

\end{tabular}

  \caption{(a) Original WSJ0 utterance ``When its initial public offering is completed Ashland is expected to retain a 46\% stake'', (b) background car noise, (c) mixture, (d) reconstructed speech using BeamformIt, (e-h) speech estimates using 32-, 8-, 4, and 1-bit DNNs and a GEV-BAN beamformer, respectively.}
  \label{fig:spectrogram_plot}
\end{figure}

Figure \ref{fig:spectrogram_plot} shows the corresponding log-spectrograms. In particular, (a) shows the original source signal, (b) the noise, and (c) the mixture, (d-h) shows the reconstructed source signals using BeamformIt and 32-, 8-, 4-, and 1-bit DNNs using a GEV-BAN beamformer, respectively. Reduced precision DNNs generate reasonable predictions compared to the single-precision baseline. BeamformIt is not able to remove the low frequency components of the car noise. The reduced-precision DNNs are able to attenuate the car noise in the background in a similar way as the 32-bit baseline DNN.\\

This is also reflected in Table~\ref{tab:experiments_SNR}, showing the SNR improvement on the test set for experiment 1 - 5. Mask-based beamformers outperform BeamformIt in all five experiments. Reducing the bit-width slightly degrades the SNR performance. However this reduces the memory footprint of the models. There is a small difference between the mask-based beamformers, i.e. GEV performs slightly better than MVDR. In general, 8-bit mask-based estimators achieve competitive SNR scores, comparable to the full precision baseline.
\begin{table}[htb!]
\caption{SNR improvement of GEV- and MVDR beamformers for various experiments conducted on WSJ0 test sets.}
\centering
  %\vspace{-5mm}
  \begin{footnotesize}
  \begin{tabular}{|c|c|c|c|c|}
  \hline
bits & experiment 1 & GEV BAN & GEV PAN & MVDR \\
\hline
   32& BeamformIt & - & - & -0.57\\
32 & DNN  &  8.09 &  8.37 &  7.36\\
8 & DNN  & 7.61 & 8.00 & 6.77\\
4 & DNN  & 4.36 & 5.81 & 4.17\\
1 & DNN  & 5.47 & 6.30 & 4.96\\ 
\hline
bits & experiment 2 & GEV BAN & GEV PAN & MVDR \\
\hline
   32& BeamformIt & - & - & -0.46\\
32 & DNN  & 8.57 & 8.76 &  7.95\\
8 & DNN  & 8.43 & 8.63 & 7.87\\
4 & DNN  & 7.50 & 8.05 & 6.21\\
1 & DNN  & 7.60 & 8.03 & 6.72\\
\hline
bits & experiment 3 & GEV BAN & GEV PAN & MVDR \\
\hline
   32& BeamformIt & - & - & -0.20\\
32 & DNN  &  11.69 &  11.96 &  10.48\\
8 & DNN  & 10.09 & 10.53 & 10.88\\
4 & DNN  & 10.29 & 10.96 & 6.65\\
1 & DNN  & 10.71 & 11.21 & 8.83\\
\hline
bits & experiment 4 & GEV BAN & GEV PAN & MVDR \\
\hline
   32& BeamformIt & - & - & -0.19\\
32 & DNN  &  8.44 &  8.72 &  7.63\\
8 & DNN  & 8.11 & 8.46 & 7.27\\
4 & DNN  & 7.01 & 7.78 & 5.49\\
1 & DNN  & 6.62 & 7.30 & 5.76\\
\hline
bits & experiment 5 & GEV BAN & GEV PAN & MVDR \\
\hline
   32& BeamformIt & - & - & 0.35\\
32 & DNN  &  12.73 &  13.15 & 10.72\\
8 & DNN  & 12.09 & 12.62 & 9.50\\
4 & DNN  & 10.26 & 11.09 & 4.60\\
1 & DNN  & 11.25 & 12.01 & 7.31\\
\hline
\end{tabular}

\label{tab:experiments_SNR}
\end{footnotesize}
\end{table}

Table~\ref{tab:experiments_ASR1} reports the word error rate (WER). We use the 6 channel data processed with 32-, 8-, 4-, and 1-bit DNNs for speech mask estimation using GEV-PAN, GEV-BAN and MVDR beamformers. Groundtruth transcriptions were generated using original WSJ0 recordings. Single-precision networks obtained the best overall WER in all experiments. In case of reduced precision networks, 8-bit DNNs produce competitive results, when using MVDR beamformers. For the 4- and 1-bit variants the performance degrades. For experiments with more than one dominant source BeamformIt fails. In general, results for single speaker scenarios (experiment 1, 2 and 4) are better.

\begin{table}[htb!]
\caption{Word error rates on enhanced speech data of WSJ0 corpus. Speech has been processed with GEV-PAN, GEV-BAN and MVDR beamformers using 32-, 8-, 4-, and 1-bit DNNs for speech mask estimation.}
\centering
  %\vspace{-5mm}
  \begin{footnotesize}
  \begin{tabular}{|c|c|c|c|c|}
\hline
bits & experiment 1 & GEV BAN & GEV PAN & MVDR \\
\hline
 32& BeamformIt & - & - & 21.38\\
32 & DNN &  9.14 &  11.71 &  9.69\\
8 & DNN & 12.13 & 16.57 & 10.62\\
4 & DNN & 22.14 & 21.89 & 15.24\\
1 & DNN & 29.97 & 40.23 & 15.07\\
\hline
bits & experiment 2 & GEV BAN & GEV PAN & MVDR \\
\hline
 32& BeamformIt & - & - & 22.77\\
32 & DNN &  8.15 &  11.10 &  9.64\\
8 & DNN & 8.89 & 10.67 & 10.17\\
4 & DNN & 12.79 & 17.56 & 12.21\\
1 & DNN & 10.88 & 15.57 & 11.20\\
\hline 
bits & experiment 3 & GEV BAN & GEV PAN & MVDR \\
\hline
 32& BeamformIt & - & - & 84.68\\
32 & DNN &  15.38 &  17.48 &  16.24\\
8 & DNN & 24.84 & 26.02 & 24.69\\
4 & DNN & 21.94 & 29.18 & 25.76\\
1 & DNN & 20.78 & 26.79 & 21.89\\
 \hline
bits & experiment 4 & GEV BAN & GEV PAN & MVDR \\
\hline
 32& BeamformIt & - & - & 22.95 \\
32 & DNN &  13.99 &  19.12 &  14.63\\
8 & DNN & 16.00 & 21.72 & 16.47\\
4 & DNN & 27.66 & 37.00 & 20.08\\
1 & DNN & 26.69 & 38.37 & 19.68\\
\hline
bits & experiment 5 & GEV BAN & GEV PAN & MVDR \\ 
\hline
 32& BeamformIt & - & - & 80.90\\
32 & DNN &  19.80 &  27.01 &  21.04\\
8 & DNN & 24.33 & 33.23 & 23.81\\
4 & DNN & 37.21 & 49.52 & 43.03\\
1 & DNN & 31.92 & 44.09 & 31.18\\
\hline
\end{tabular}
\label{tab:experiments_ASR1}
\end{footnotesize}
\end{table}

\section{Conclusion}
\label{sec:con}
We introduced a resource-efficient approach for multi-channel speech enhancement using DNNs for speech mask estimation.
In particular, we reduce the precision to 8-, 4- and 1-bit. We use a recurrent neural network structure capable of learning long-term relations. 
%By successively we provide an resource-efficient approach for multi-channel speech enhancement based on Deep Neural Networks (DNNs).
Limiting the bit-width of the DNNs reduces the memory footprint and improves the computational efficiency while the degradation in speech mask estimation performance is marginal. When deploying the DNN in speech processing front-ends only the reduced-precision weights and forward-pass calculations are required. This supports speech enhancement on low-cost, low-power and limited-resource front-end hardware.
We conducted five experiments simulating various cocktail party scenarios using the WSJ0 corpus. In particular, different beamforming architectures, i.e.\ MVDR, GEV-BAN, and GEV-PAN, which are combined with low bit-width mask estimators have been evaluated. MVDR beamformers, using 8-bit reduced-precision DNNs for  estimating the  speech mask, obtain competitive SNR scores compared to the single-precision baselines. Furthermore, the same architecture achieve competitive WERs in single speaker scenarios, measured with the Google Speech-to-Text API. If multiple speakers are introduced, the performance degrades. In the case of binary DNNs, we show a significant reduction of memory footprint while still obtaining an audio quality which is only slightly lower compared to single-precision DNNs. We show that these trade-offs can be readily exploited on today's hardware, by benchmarking the core operation of binary DNNs on NVIDIA and ARM architectures.

In future, we aim to implement the system on a target hardware and measure the resource consumption and run time.

%\input{eigennet_architectures}
%\input{experiments_wsj0}
%\input{experiments_chime4}
%\newpage
%\input{conclusion}


\bibliographystyle{IEEEtran}
\bibliography{bibliography,nnvibinary}


\end{document}